\documentclass[final,5p,twocolumn]{elsarticle}

\usepackage{amssymb}
\usepackage[normalem]{ulem}
\usepackage{color}
\usepackage{epstopdf}
\usepackage{graphicx}
\usepackage{flushend}

\journal{Cement and Concrete Research}

\begin{document}
	
	\begin{frontmatter}
		
		
		
		\title{Rheology of  hydrating cement paste: crossover between two aging processes}
		
		
		\author[lab1,lab2]{Atul Varshney\corref{cor1}}
		\ead{atul.varshney@weizmann.ac.il}
		\address[lab1]{Department of Condensed Matter Physics and Materials Science, Tata Institute of Fundamental Research, Homi Bhabha Road, Mumbai 400 005, India}
		\address[lab2]{Department of Physics of Complex Systems, Weizmann Institute of Science, Rehovot, Israel 76100}
		
		\author[lab1]{Smita Gohil\corref{cor1}}
		\ead{smitagohil@tifr.res.in}
		\author[lab1]{B. A. Chalke}
		\author[lab1]{R. D. Bapat}
		\author[lab3]{S. Mazumder}
		\address[lab3]{Solid State Physics Division, Bhabha Atomic Research Centre, Mumbai 400 085, India}
		\author[lab1]{S. Bhattacharya}
		\author[lab1]{Shankar Ghosh\corref{cor1}}
		\ead{sghosh@tifr.res.in}
		\cortext[cor1]{Corresponding author}
\begin{abstract}
The roles of applied strain and temperature on the hydration dynamics of cement paste are uncovered in the present study. We find that the system hardens over time through two different aging processes. The first process dominates the initial period of hydration and is characterized by the shear stress $\sigma$ varying sub-linearly with the strain-rate $\dot{\gamma}$; during this process the system is in a relatively low-density state and the inter-particle interactions are dominated by hydrodynamic lubrication. At a later stage of hydration the system evolves to a high-density state where the interactions become frictional, and $\sigma$ varies super-linearly with $\dot{\gamma}$; this is identified as the second process. An instability, indicated by a drop in $\sigma$, that is non-monotonic with $\dot{\gamma}$ and can be tuned by temperature, separates the two processes. Both from rheology and microscopy studies we establish that the observed instability is related to fracture mechanics of space-filling structure.
		\end{abstract}
		
		\begin{keyword}
			Cement paste \sep Hydration \sep Rheology \sep Microstructure  \sep Aging
		\end{keyword}
		
	\end{frontmatter}
	
	
	\section{Introduction}
	\label{Introduction}
	Hardening process of cement in the presence of water has been a subject of research besides its industrial importance \cite{roussel_understanding_2012}. Cement paste, usually a dense suspension of non-Brownian particles, exhibits complex rheological properties that depend on several factors including particles' shape and size, cement composition, cement$/$water ratio, measurement methods etc \cite{larson_structure_1999, lapasin_1979, lootens_2004}. In the presence of water, ordinary portland cement (OPC)$-$ a commonly used binding agent$-$ forms gel-like structure that arises predominantly from the chemical reactions producing calcium-silicate-hydrates (C-S-H) \cite{yammine_2008, schmidt_rheological_2002, h_f_w_taylor_cement_1997}. Besides C-S-H, aluminates$-$ a fast reacting phase present in OPC$-$ form space filling needle-like structures called Ettringrites. This is accompanied with the growth of a  cohesive system-spanning network of C-S-H which bridges the hydrating cement grains. As a result  stress-bearing submicron structures form in the system which offer   resistance in response to applied  shear. The Ettringites, however, contribute to the  solidification  of cement only during its initial stage of hydration \cite{roussel_understanding_2012}.
	
	As the hydration reaction proceeds over time, the  morphology of the C-S-H network grows in a manner that gives rise to an overall decrease of pore density and relatively a compact structure evolves \cite{lootens_2004, beaudoin_1994}. With the availability of water running out the interparticle interactions  in this stage of hydration are mainly  frictional in nature \cite{yammine_2008, gallucci_2013}. Hardening of the system at long times is analogous to the ubiquitous phenomena of aging observed in frictional systems \cite{ruina_slip_1983}. It relates to the reconstruction of the interfacial contact zones \cite{ruina_slip_1983, li_frictional_2011} and flow induced reconfigurations of its constituents, such that, a low density cluster of particles which has fewer contacts with its neighbors evolves into a denser structure that shares more contacts with its neighbors \cite{brown_shear_2014, bai_adiabatic_1992, langer_shear-transformation-zone_2008}. 
	
	Aggregation and network formation are   out-of-equilibrium phenomena that are commonly influenced by mechanical perturbations, e.g., shear flow \cite{lapasin_1979, roussel_origins_2012, roussel_rheology_2007}. Such  perturbations  would lead to mechanical failure \cite{sollich_rheology_1997} at multiple length scales., e.g., at the scale of the contact region between the particles it is related predominantly to the breaking of the calcium-silicate-hydrate bonds while at larger scale it relates to the shear-induced transformation of  particle configurations \cite{roussel_origins_2012}. However, most studies that address the issues related to the emergence of  mechanical strength in cement paste have been performed in the presence of small external stresses that perturb the system about a local minima of its potential energy landscape  \cite{roussel_understanding_2012, roussel_rheology_2007, bellotto_2013, nachbaur_2001}.
	
	{Temperature plays multiple roles; it enhances the rate of hydration \cite{gallucci_2013}, determines the filling of the interstitial space between the grains and allows thermally activated processes to release the internal stresses {\it via} creep \cite{Elkhadiri_2009, coussot_avalanche_2002}. Arguments in similar lines are usually proposed to understand the variation of the mechanical properties of cement on its curing temperature \cite{Elkhadiri_2009, elkhadiri_effect_2008}. Cement cured at low temperatures develops a more compact structure with improved mechanical stability  through a slow and controlled process of filling up the interstitial space between the cement grains. In contrast, curing at high temperature produces non-uniform structure with low compressive strengths \cite{elkhadiri_effect_2008, beaudoin_pore_1994}.

	In this paper we study the hardening process of cement paste as a function of temperature and  in presence of large-scale mechanical perturbations (large oscillatory shear) where the imposed shear disrupts the energy landscape by constantly restructuring the interwoven matrix of hydrate needles. We observe that the cement paste gains shear rigidity via two distinct aging processes; the first  aging process is associated with the \textit{setting} process  that generates space-filling structures (e.g.,  Ettringite needles) and  provides the system its initial rigidity.  This system has  flow curves that show shear-thinning behavior. In contrast, the second  process is associated with the \textit{hardening}  phenomena that  arises from the formation of C-S-H networks and the material manifests shear-thickening behavior.  In addition, the crossover from the first to the second aging process occurs through a global weakening in the system,  indicated by a drop in the measured shear stress. We also show that with lowering temperature the drop occurs later in time and its magnitude decreases. Both these phenomena  are related to the reaction kinetics of the hydrates formation. Finally, we establish that the weakening is related to the fracture mechanics of the  space-filling structure formed during the initial phase of cement hydration. 
\section{Experiment}
In experiments, we use a cement paste prepared by mixing OPC ($c$) with deionized water ($\mbox{w}$) in  a ratio $\mbox{w}/c=0.5$ for about a minute and then immediately transferred to the rheometer for measurements \cite{mazumder_temporal_2005}. This  water to OPC ratio  allows us to  access the  hydrodynamic lubrication regime  between the cement grains during its initial phase of hydration. We also perform the rheological measurements on Alite paste, prepared in the same water to cement ratio, and compare the data with that of  OPC.  Alite ($\mbox{C}_3\mbox{S}$; tricalcium silicate)$-$ a major phase present in the cement$-$ is believed to contribute primarily to the mechanical strength of hydrating cement paste through formation of C-S-H; it does not produce Ettringites \cite{h_f_w_taylor_cement_1997,  nachbaur_dynamic_2001, barnes_2002}.
\subsection{Microscopy}
The scanning electron micrographs in Fig. \ref{fig1} capture the evolution in the morphology of the cement particles during the process of hydration at temperature $T=30^{\circ}\mbox{C}$. 
In the early stage of hydration, Ettringite (calcium sulfoaluminate hydrate) forms, which is a faster growing compound and can be seen as needle-shaped structures in Fig. \ref{fig1}(b-d), and primarily occupies more volume in the system.
\begin{figure}[t!]	
\centering
\includegraphics[scale=0.295]{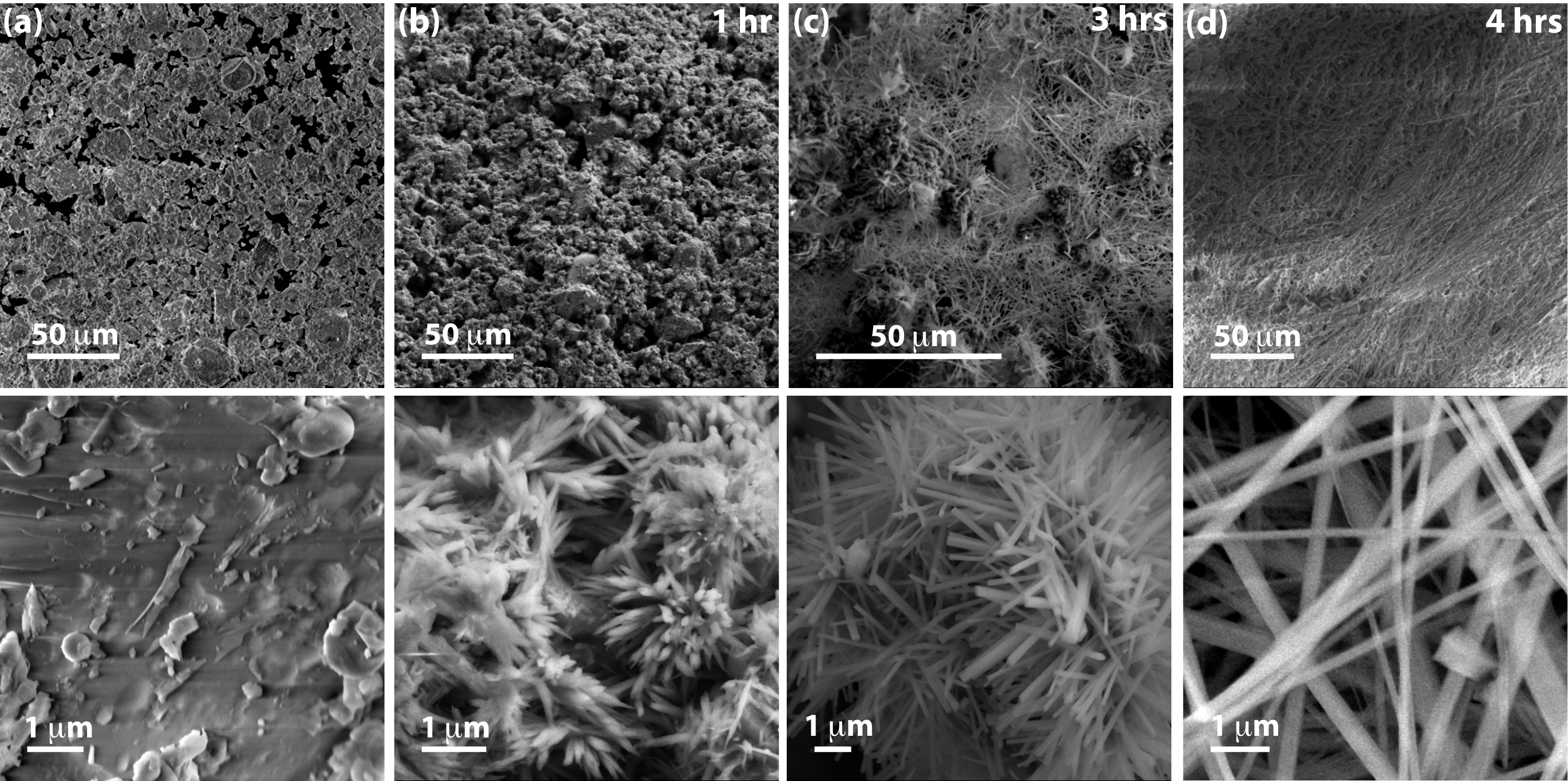}
\caption{(a) Scanning electron micrographs of dry cement  powder grains. (b-d) Low  (top panel) and high  (bottom panel) magnification micrographs of hydrated cement at $30^{\circ}\mbox{C}$ with $t=1~\mbox{hr}$, $3~\mbox{hrs}$ and  $4~\mbox{hrs}$, respectively. The interweaved needle structure and its growth with time are evident in the high magnification images.}
\label{fig1}
\end{figure}
\subsection{Oscillatory Rheology} 
A conventional strain-controlled rheometer (MCR301;  Anton Paar) is used to perform oscillatory rheological measurements of the cement paste as a function of time $t$ and at various temperatures $T$. The cement paste is subjected to a sinusoidal shear strain of amplitude $\gamma$  at a fixed oscillation frequency $\omega$ and correspondingly waveform of torque  is being recorded by the rheometer. Typically the time varying angular displacement  of the measuring device and the measured torque are converted  to strain  and shear stress, respectively by multiplying them with suitable `form factors' which are specific to a given measuring system. However, the conversion from angular displacement to strain is obtained by assuming a linear velocity profile in the measurement gap region and no-slip conditions at the walls. For some applications this  assumption may not  strictly hold, e.g., avalanche motion in sand piles is  a surface phenomenon where   the  particulate matter  screens the stress  such that only a shallow  region  close to the moving surface is sheared \cite{atman_stress_2005}. For these systems both  shear stress and  strain are underestimated by the rheometer \cite{hyun_review_2011}. 

The measurements are done in two methods: (1)  a parallel-plate geometry is used (see schematic in Fig. \ref{fig1_schematic}(a)) and oscillatory strain of amplitude $\gamma=1\%$  at  $\omega=6.28~\mbox{rad/s}$  is applied to the cement paste. The process of hardening of the cement paste is monitored by measuring the amplitude $\sigma$ of the waveform associated to the shear stress as a function of hydration time ($t$).  The measurement is repeated on fresh cement paste for different gap height $h$ between the parallel plates  at $T=30^\circ\mbox{C}$, and for different $T$ at a fixed gap $h=2~\mbox{mm}$. Grounded glass plate (radius $R=25~\mbox{mm}$) is used as the shearing surface to ensure sustained contact with the cement paste. The strain amplitude ($\gamma$) is measured at the circumference of the top plate through $\gamma=R\phi/h$, where $\phi$ is the angular displacement. (2) a concentric cone-cylinder geometry (cone radius $r_1=13.3~\mbox{mm}$, cylinder radius $r_2=14.5~\mbox{mm}$, cone angle=$120^{\circ}$) is used to measure a set of flow curves, i.e., $\sigma$ versus $\dot\gamma$, at $\omega=1~\mbox{rad/s}$ and at time intervals of $1~\mbox{hr}$ while the system is maintained at a constant $T$. The  strain-rate ($\dot{\gamma}$) in this geometry is equal to the amplitude of $(\frac{r_2^2+r_1^2}{r_2^2-r_1^2})(\frac{d\phi}{dt})$. A schematic of the cone-cylinder geometry is shown in Fig. \ref{fig1_schematic}(b).

\section{Results and discussion}

In this section we first report our experimental observations in the  parallel-plate geometry and  later provide a detailed account of observed phenomena  in the cone-cylinder geometry.  
\subsection{Parallel-plate geometry} 
\subsubsection*{(a) Temporal evolution of shear stress} 

	 	\begin{figure}[t!]
		\centering
		\includegraphics[scale=0.3]{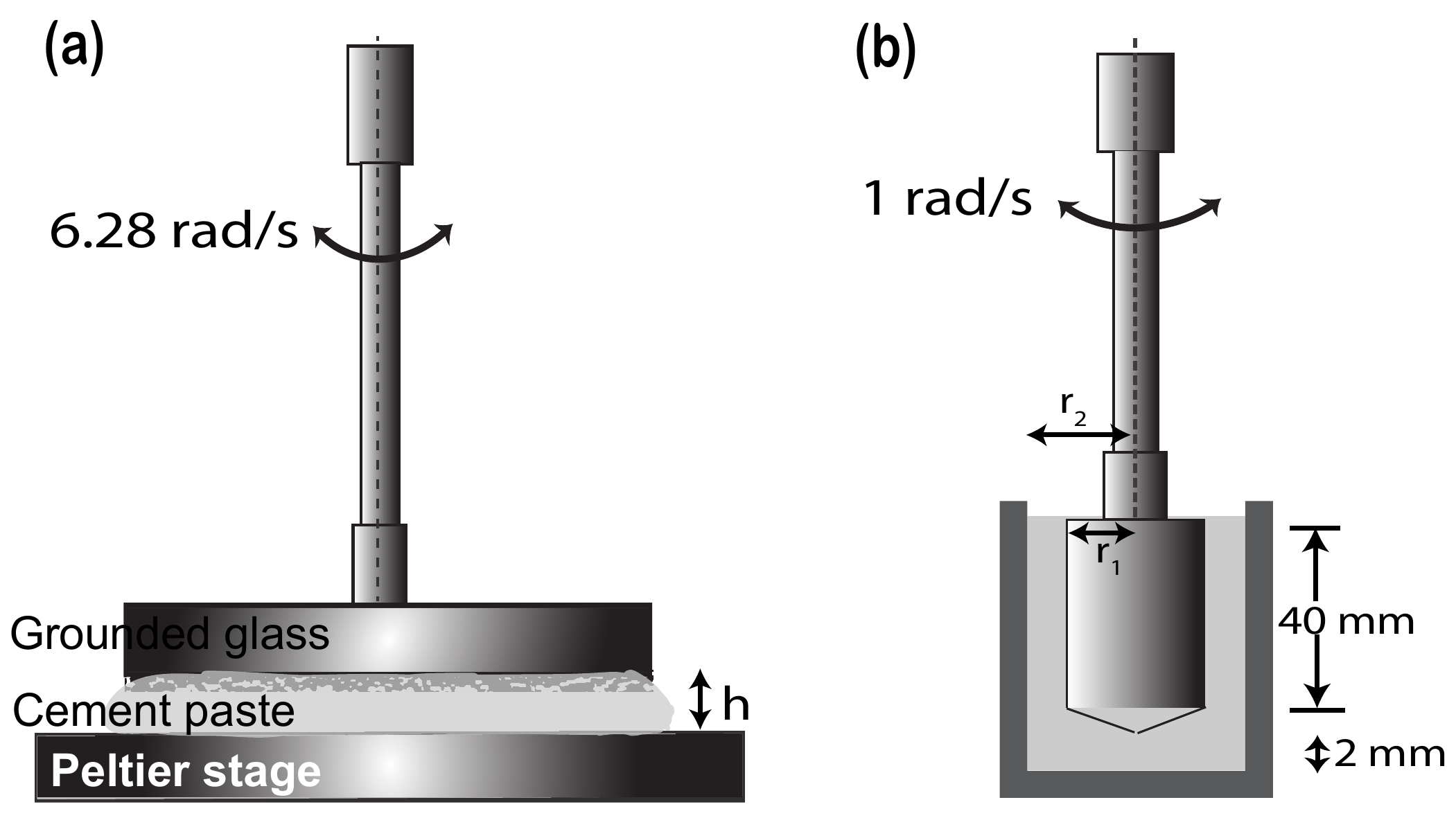}
		\caption{Schematics of measurement geometries (not to scale): (a) parallel-plate (b) concentric cone-cylinder. }
		\label{fig1_schematic}
	\end{figure}

Figure \ref{fig2}(a) shows a typical variation of $\sigma$ with $t$, obtained for $\gamma=1\%$ and at $T=30 ^{\circ}\mbox{C}$. The increase in $\sigma$ above $t_0$ is  followed by a pronounced drop ($\Delta\sigma$) at $t =t^{\ast}$ (and $\sigma=\sigma^{\ast}$).  It suggests that the system begins to weaken at $t=t^{\ast}$ and  regains its strength later. The weakening of the structure can come about by the  initiation of  bulk flows in the system. Such flows  would result in a large-scale reconfiguration  via the  alteration of  the  space-filling matrix brought by the applied shear  \cite{roussel_steady_2005}.  In our experiments, we find $\gamma=0.1\%$ is not effective  to disrupt the structure hence $\sigma$ grows with $t$ at $T=30 ^{\circ}\mbox{C}$, shown in the inset of Fig. \ref{fig2}(a), which is in agreement with results reported in ref. \cite{bellotto_2013, nachbaur_dynamic_2001}.  As the hydration reaction proceeds, the density of cohesive C-S-H network increases \cite{gallucci_2013}, as a result $\sigma$ regains, but with a faster rate. 

The above observations are corroborated by the measurement of the phase angle ($\delta$) between waveforms of shear stress and strain. For viscoelastic fluids, $\delta$ varies between 0  (purely elastic response) and  $\pi/2$ (purely viscous response) \cite{larson_structure_1999}.  Figure \ref{fig2}(b) shows the time evolution of $\tan\delta$ that increases till the point system weakens, thus it is suggestive of enhanced viscous response of the system. As the system regains its strength, $\tan\delta$ decreases with $t$, which indicates the subsequent growth of load bearing percolating C-S-H networks in the system. Strikingly, the value of $\tan\delta$ at long times is considerably smaller than its value at initial times which signifies the growing solid-like behavior in the system (see also Fig. \ref{fig2}(b) inset for different $\gamma$).

The advantage of using parallel-plate geometry is that the gap between the two plates can be adjusted. We find that the drop in shear stress ($\Delta\sigma$) normalized with  its peak value ($\sigma^{\ast}$) increases initially with the gap height $h$ and then saturates to a value close to unity, as shown in Fig. \ref{fig2}(c). This shows that the drop observed in $\sigma$ (Fig. \ref{fig2}(a)) is due to the volumetric disintegration of the structures.   To verify the occurrence of the stress drop, rheological measurements are performed on Alite paste at $T=18 ^{\circ}\mbox{C}$. We find that $\Delta\sigma$ for Alite is about an order of magnitude smaller than that for OPC, as shown in the inset of Fig. \ref{fig3}(a). Thus  
the weakening of the system, i.e., the presence of $\Delta\sigma$ in OPC, arises due to the disruption of the  space-filling structure.

	\begin{figure}[t]
		\centering
		\includegraphics[scale=0.305]{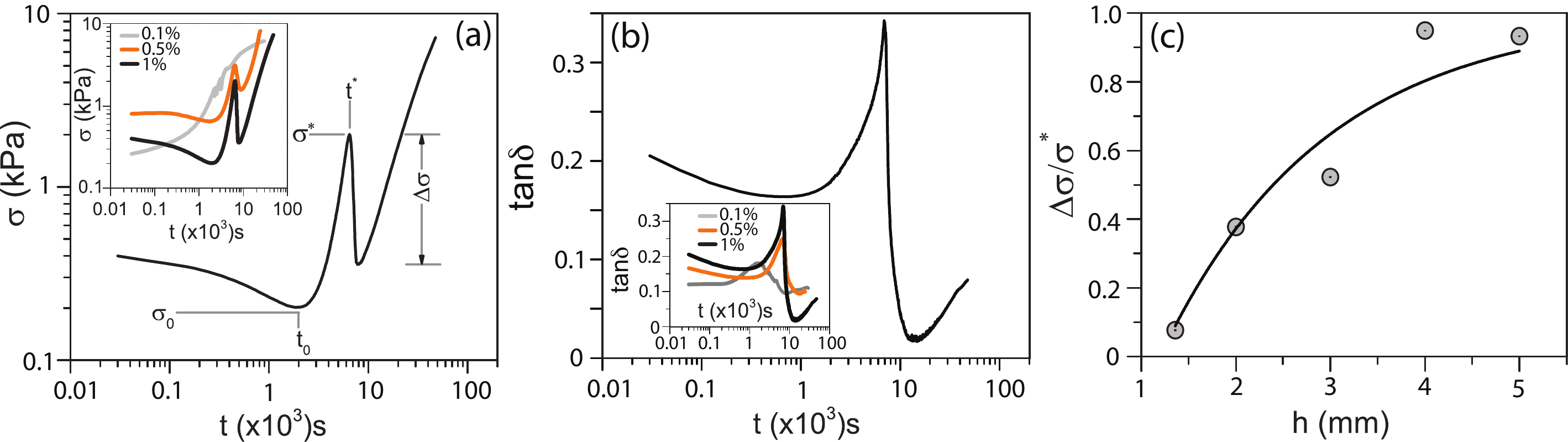}
		\caption{Variations of $\sigma$  and $\tan\delta$  with hydration time $(t)$ at $T=30 ^{\circ}\mbox{C}$ and $\gamma=1\%$ are plotted in (a) and (b), respectively. The insets to (a) and (b) compare  the time variation of $ \sigma $ and $ \tan \delta $, respectively  for $\gamma=0.1\%$ (gray), $0.5\%$ (orange) and $1\%$ (black), at $T=30 ^{\circ}\mbox{C}$. (c) Variation of $\Delta \sigma$ normalized with $\sigma^{\ast}$  for different gaps $h$. Solid line is a guide to the eye. }
		\label{fig2}
	\end{figure}
	\begin{figure}[t]
		\centering
		\includegraphics[scale=0.4]{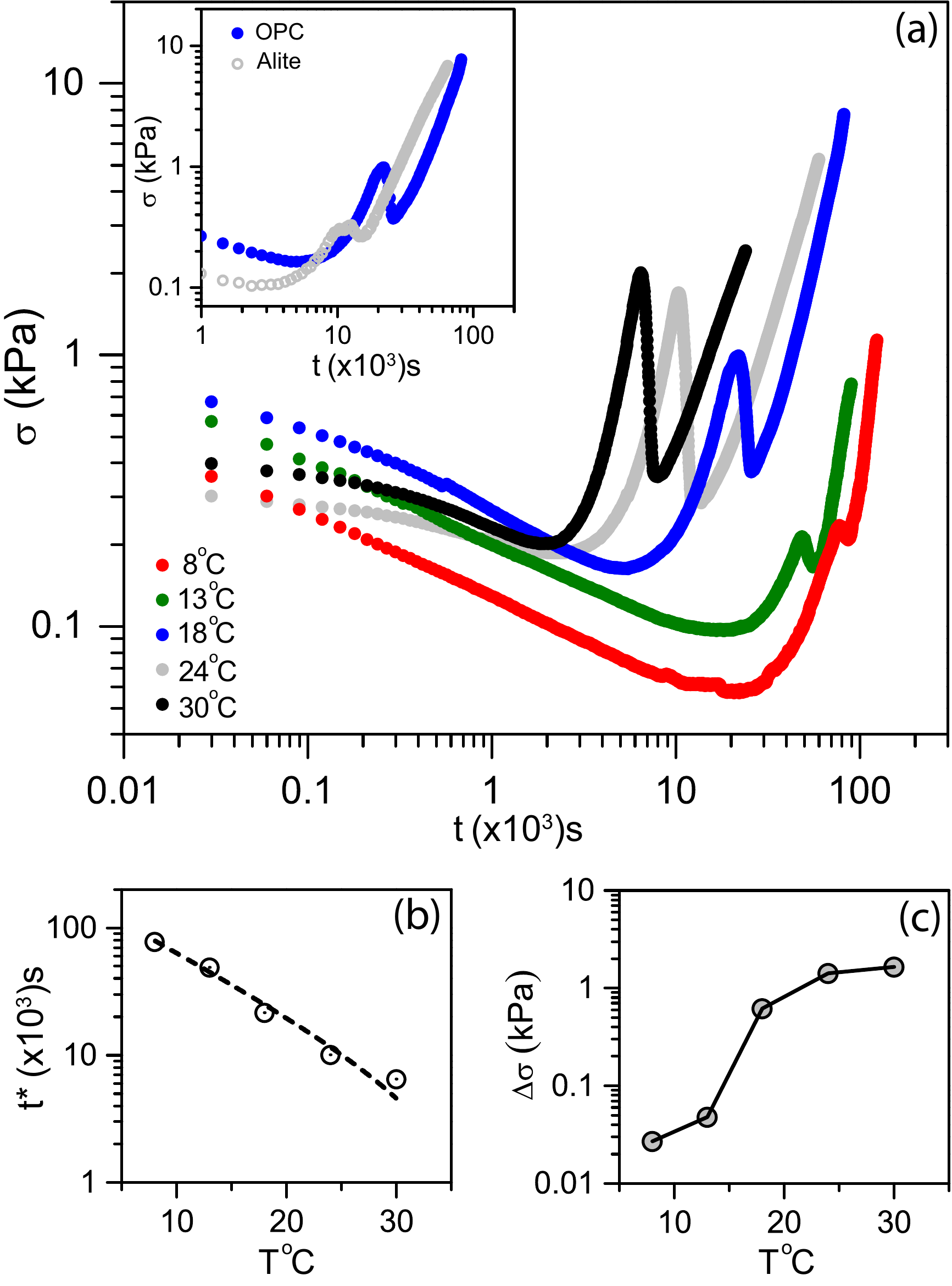}
		\caption{(a) Time evolution of $\sigma$ at $\gamma=1\%$ and for different $T$. Inset shows the variation of $\sigma$ with $t$ for Alite (gray) and OPC (blue) measured at $18^{\circ}C$. (b) Variation of $t^*$  with $T$;  dashed line is an exponential fit to the data. (c) Stress drop ($\Delta \sigma $) versus $T$.}
		\label{fig3}
	\end{figure}
	\subsection*{(b) Role of temperature}
	To further study the dependency of stress drop on the rate of structure formation we perform the rheological measurements at different constant temperatures. Figure \ref{fig3}(a) demonstrates the time evolution of $\sigma$ for various $T$. The ensuing growth of $\sigma$ with $t$ is not only delayed with lowering of temperature, its behavior also changes qualitatively. The drop in the stress ($ \Delta \sigma $) gets progressively less pronounced with lowering of temperature (see Fig. \ref{fig3}(a)). The  hydration rate  determines the peak position, i.e., $t^{\ast}$,  which occurs later in time with  decreasing $T$ (see Fig. \ref{fig3}(b)). A faster rate of growth would indeed produce a higher density of hydrates over a given period of time and vice versa \cite{gallucci_2013}. The drop in $\sigma$, proportional to the number of failure events, is related to the density of  hydrates and hence $\Delta\sigma$ is expected to increase with $T$. This is indeed the case  shown in Fig. \ref{fig3}(c). 
	
	\begin{figure}[t]
		\centering
		\includegraphics[scale=0.38]{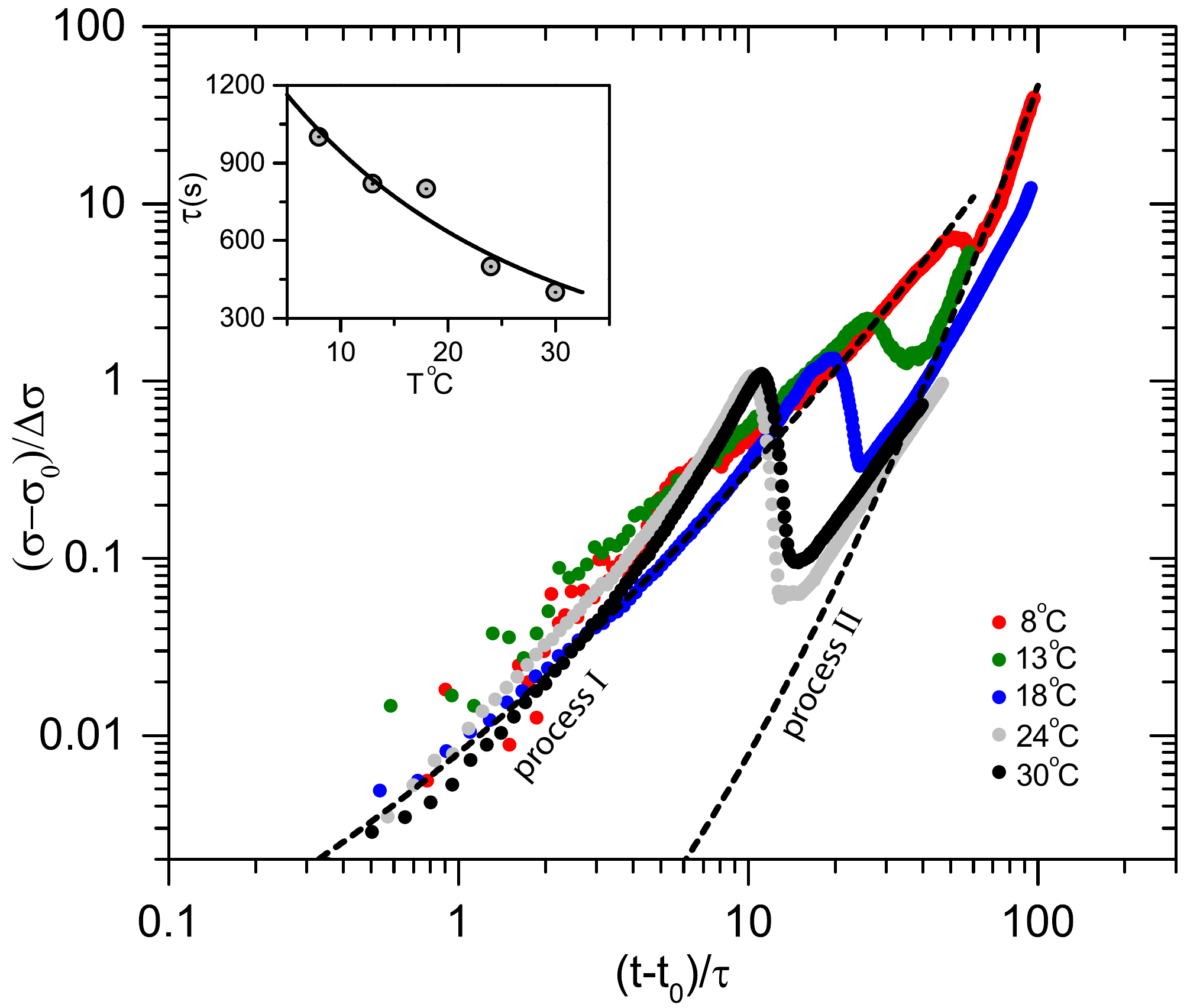}
		\caption{Scaling plot, ($\sigma-\sigma_0)/\Delta \sigma$ with ($t-t_0)/\tau$ for various $T$,  shows the existence of  two  processes (highlighted by dashed lines) of aging. Here $\tau$ is a scaling parameter whose variation with $T$ is shown in the inset. The solid line is an exponential fit of the form $\tau\sim e^{-U/k_B(273+T)}$, which yields $U\sim 0.5~\mbox{eV}$.}
		\label{fig4}
	\end{figure}
	
	\subsection*{(c) Evidence of two aging processes }The difference in the asymptotic variations of $\sigma$ with $T$ for $t_0<t<t^{\ast}$ and $t\gg t^{\ast}$ is illustrated in Fig. \ref{fig4}. Here we scale the various $\sigma-T$ graphs in the region $t>t_0$. The scaling is obtained by shifting the origin of the plots to ($\sigma_0,~t_0$) and then scaling the ordinate, i.e., ($\sigma-\sigma_0$) by $\Delta\sigma$ and the abscissa, i.e., ($t-t_0$) by a scaling parameter $\tau$. The temperature variation of $\tau$ is shown in the inset of Fig. \ref{fig4}. In a manner similar to most thermally activated processes \cite{ediger_supercooled_1996}, $\tau$ decays exponentially with $T$, and one obtains an energy scale $U\sim0.5~\mbox{eV}$ after fitting the variation to the Arrhenius form $\tau\sim e^{-U/k_B(273+T)}$, where $k_B$ is the Boltzmann constant.
	
The scaling graph illustrates the existence of two processes, shown by dashed lines in Fig. \ref{fig4}, via which the aging  proceeds. The weakening is related to the crossover between the two  processes. For smaller values of $(t-t_0)/\tau$, the rate of aging is faster in  process \textbf{II} as compared to  process \textbf{I}. However, the aging rate for the  process \textbf{I} catches up with  process \textbf{II} only later and thus the magnitude of weakening gets smaller further along the abscissa. In the context of cement,  \textbf{I}  and \textbf{II} correspond to the \textit{setting} and \textit{hardening} processes, respectively \cite{nachbaur_2001}.

	\subsection{Cone-cylinder geometry} 
\subsection*{Flow curves: crossover from shear thinning to thickening}
\begin{figure}[t]
		\centering
		\includegraphics[scale=0.4]{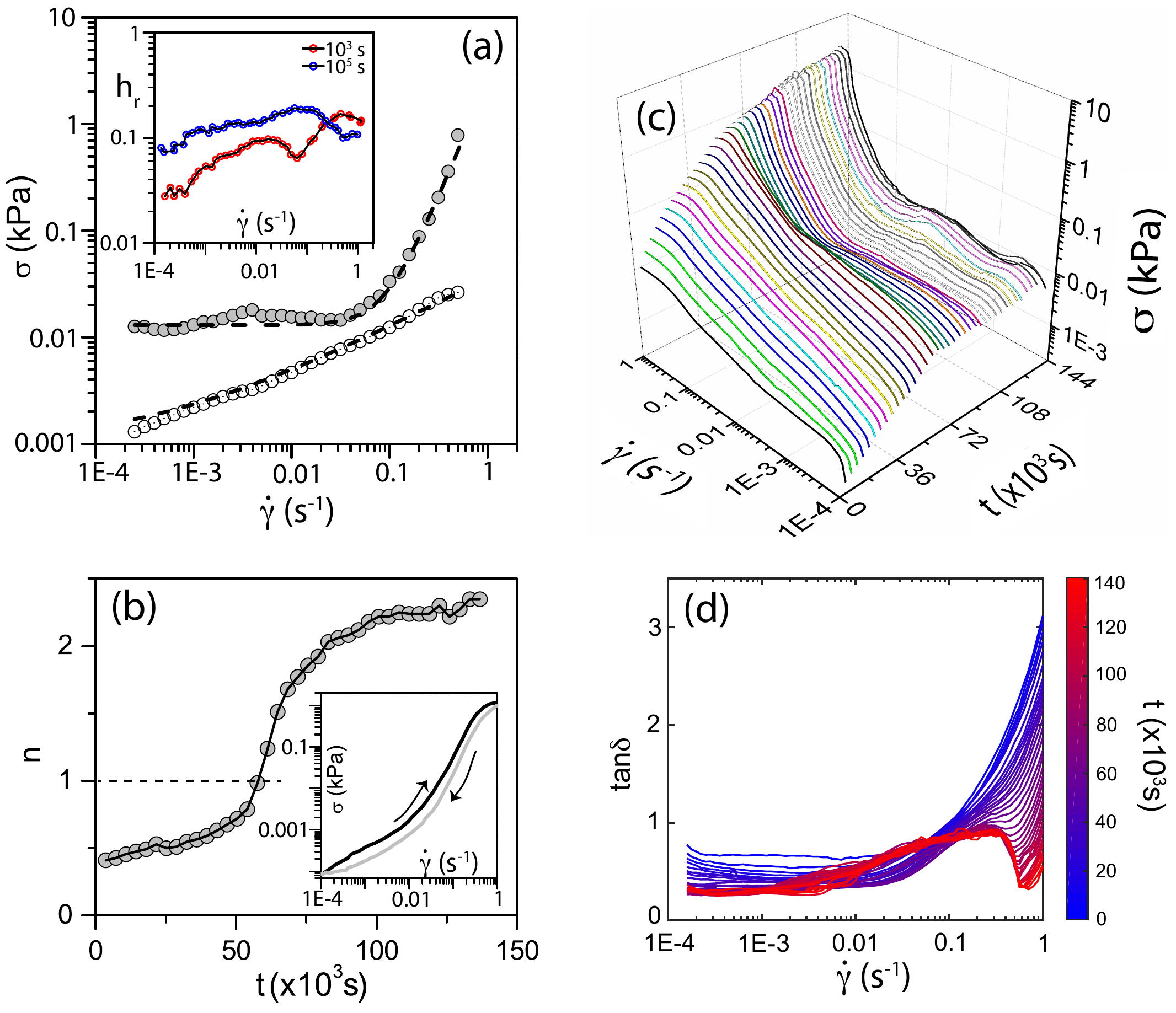}
		\caption{(a) Representative flow curves, $\sigma$  versus $\dot\gamma$, at $\omega=1~\mbox{rad/s}$ and $T=8^{\circ}\mbox{C}$, for $t=10^3\mbox{s}$ (open circles) and  $t=10^5\mbox{s}$ (filled circles). The dashed lines are fit to the Herschel-Bulkley form. The fit yields $n=0.47$, $\sigma_Y =1~\mbox{kPa}$, $A=4~\mbox{kPa}\cdot \mbox{s}^{0.47}$ for $t=10^3\mbox{s}$ and $n =2.2$, $\sigma_Y=13~\mbox{kPa}$,  $A=0.1~\mbox{kPa}\cdot \mbox{s}^{2.2}$ for $t=10^5\mbox{s}$. The inset shows variation of $h_r$ with $\dot\gamma$ for two typical waveforms  at $t= 10^3\mbox{s}$ (red) and $t=10^5\mbox{s}$ (blue). (b) Variation of exponent $n$ with  $t$.  Inset shows a typical $\sigma-\dot\gamma$ hysteresis curve obtained at  $t=10^5\mbox{s}$.  (c) Stacked plot of $\sigma-\dot\gamma$ curves obtained at $T=8^{\circ}\mbox{C}$; each curve is recorded at an interval of 1 hour. (d) Plot of $\tan\delta-\dot\gamma$ curves obtained at $T=8^{\circ}\mbox{C}$; each curve is recorded at an interval of $1~\mbox{hr}$. The colormap shown beside displays the hydration time $t$;  blue signifies the early stage of hydration and red to the later times.}
		\label{fig5}
	\end{figure}
	To explore further the crossover from  process \textbf{I} to  process \textbf{II}, we record flow curves, $\sigma$ versus $\dot\gamma$ at a frequency  $\omega=1~\mbox{rad/s}$  and at time intervals of one hour in the cone-cylinder geometry at $T=8 ^{\circ}\mbox{C}$.  Two representative flow curves for $t=10^3\mbox{s}$ (open circles) and $t=10^5\mbox{s}$ (filled circles)  are shown in Fig. \ref{fig5}(a).  The flow curves  fit well to the  general Herschel-Bulkley (HB) form \cite{roussel_understanding_2012,larrard_fresh_1998}, $\sigma=\sigma_Y+A{\dot{\gamma}}^n$, where $\sigma_Y$ is the yield stress, strain-rate $\dot{\gamma}=\omega\gamma$, $A$ and $n$ are additional fitting parameters, see Fig. \ref{fig5}(a). For $n=1$, the material behaves like a Bingham plastic that flows like a Newtonian liquid ($\sigma\propto\dot{\gamma}$) above $\sigma_Y$ \cite{otsubo_1980}. For $n<1$ and $n>1$ the system softens or stiffens, i.e., offers lower or higher differential resistance to flow with increasing $\dot{\gamma}$, respectively.  With time the value of $n$ crosses from $n<1$ (shear-thinning) to $n>1$ (shear-thickening), see Fig. \ref{fig5}(b).  The crossover occurs at $t\sim 60\times10^3\mbox{s}$ and it is coincident with the drop in $\sigma$ for $T=8^{\circ}\mbox{C}$ (see Fig. \ref{fig3}(a)). It is also reflected in the time evolution of  stacked flow curves, plotted in Fig. \ref{fig5}(c) and in the variation of $\tan\delta$ with $\dot\gamma$ for different $t$ (Fig. \ref{fig5}(d)).

The  crossover from shear thinning to thickening  is also accompanied by qualitative change in the nature of the $\tan\delta-\dot\gamma$ curves (Fig. \ref{fig5}(d)). At early stages of hydration, the variation of $\tan\delta$ remains flat for low $\dot\gamma$ and increases with $\dot\gamma$ for $\dot\gamma\gtrsim0.02~\mbox{s}^{-1}$. The increase in $\tan\delta$ is related to the enhanced viscous dissipation in the system  that arises due to the strain-induced flow.  However, at later stages of hydration and beyond $\dot\gamma\sim0.3~\mbox{s}^{-1}$,  $\tan\delta$ decreases with $\dot\gamma$ which is an indication of growing solid-like response of the system. In this stage, transient load bearing C-S-H structures form stochastically in the system and the interparticle interactions evolve to frictional type \cite{yammine_2008}. Thus, we interpret the crossover of $n$ (Fig. \ref{fig5}(b)) as a signature of the time evolution of inter-particle interactions \cite{schmidt_rheological_2002, mahaut_effect_2008}. For small times, it is of the lubrication type formed by the presence of water and shear-induced fluidized hydrates in the interstitial spaces between the cement particles, however, at later times with the water being converted into hydration products the interparticle interactions are mainly dominated by  frictional interactions \cite{yammine_2008}.

During the process of obtaining the flow curves, the system is subjected to increasing values of strain which causes failure of  space-filling and C-S-H structures. Thus, the flow properties of the system at any  point during the experiment is determined by the  competitive dynamics of the strain-induced  mechanical  failures  and the hydration-induced growth of the load bearing C-S-H networks in the system. Further, during each flow curve the system is subjected to a large extent of deformation, i.e.,  $\gamma \sim 100\%$ or $\dot\gamma\sim1~\mbox{s}^{-1}$. Therefore, the resulting flow erases all memory of earlier configurations and produces a fresh state of the system at the end of each strain cycle \cite{roussel_origins_2012, struik_rejuvenation_1997, fourmentin_rheology_2015}. Moreover, at the macroscopic scale the system is almost reversible under strain cycling which is evident from a minor hysteresis loop of $\sigma-\dot\gamma$, shown in the inset of Fig.  \ref{fig5}(b). 
	
	The nonlinearity in the measurements is quantified by computing the total harmonic distortion ($h_r$)  from the recorded torque  waveforms \cite{hyun_review_2011}.  The total harmonic distortion is defined as $h_r=\sqrt{(M_2^2+M_3^2+\ldots)}/M_1$; where $M_1, M_2, M_3,\ldots$ are the amplitudes of first, second, third,$\ldots$ harmonic components of $M$, respectively \cite{shmilovitz_2005}.  In our experiments, we considered up to seven harmonics. The inset of Fig. \ref{fig5}(a) shows the variation of $h_r$ with $\dot\gamma$ for $t=10^3\mbox{s}$ and $10^5\mbox{s}$.    The overall distortion increases as the system evolves in time from shear thinning  to thickening type, however their variation with $\dot\gamma$ remain almost flat (Fig. \ref{fig5}(a) inset).

\subsection{Microscopic evidence of disruption of space-filling structure}
Our   assertion  by comparing the time evolution of $\sigma$ for Alite and OPC  (see Fig. \ref{fig3}(a) inset) that $\Delta\sigma$ is associated with the fracture mechanics of the space-filling structure is  supported by micrographs obtained for cement paste that has been sheared for $\sim4~\mbox{hrs}$ at $30^{\circ}\mbox{C}$  shown in  Fig. \ref{fig7}. The structures seen in the sheared paste are smaller than that observed for similarly aged unsheared paste, see Fig. \ref{fig1}(d).  
\begin{figure}[t]
\centering
\includegraphics[scale=0.38]{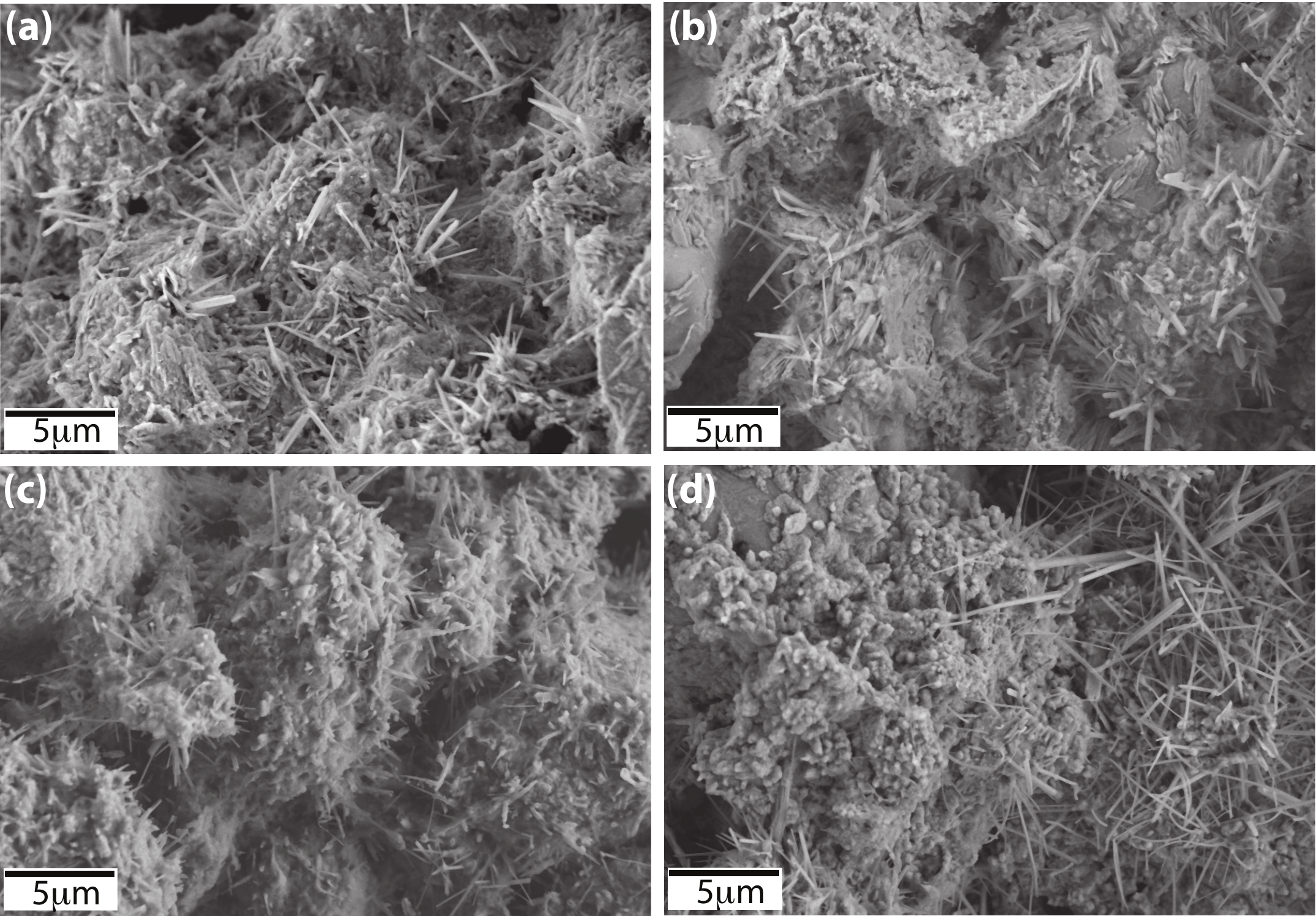}
\caption{(a-d) Scanning electron micrographs of cement paste sheared at $30^{\circ}\mbox{C}$ for $\sim4~\mbox{hrs}$ at different gaps $h=1.3~\mbox{mm}$, $2~\mbox{mm}$, $4~\mbox{mm}$ and $5~\mbox{mm}$, respectively. To compare with similarly aged unsheared paste, see Fig. \ref{fig1}(d). }
\label{fig7}
\end{figure}

\section{Conclusions}
In conclusion, we present results from rheology of cement paste when subjected to a large amplitude shear: 
	(\textbf{i}) time evolution under constant shear shows a drop in shear stress, i.e., $\Delta\sigma$, for $\gamma\ge$ 0.5$\%$. (\textbf{ii}) lowering of temperature  delays the onset of growth of $\sigma$;  $\Delta\sigma$ becomes less pronounced. (\textbf{iii}) two  processes are identified {\it{via}} which the aging proceeds;  process \textbf{I} corresponds to the lubrication dominated regime and  process \textbf{II} to the friction dominated one. The weakening, $\Delta\sigma$, is related to the crossover between the  processes. (\textbf{iv}) the flow curves in  process \textbf{I} reveal the shear-thinning behavior of material while those in process \textbf{II} show shear-thickening. (\textbf{v}) large strains produce long-range particle rearrangements and as a consequence disrupts the fabric nature of the space-filling structure, as seen from the microscopy results of the cement paste that has been sheared for several hours.

	Thus, we show that mechanical perturbations irrevocably alter the interwoven fabric of space-filling structure, resulting in weakening of the system. The transition {\it{via}} weakening between the two observed  processes bears similarities to the structural phase transitions identified in amorphous materials and  are related to local plastic deformations that often condensate to form a global soft mode \cite{loerting_multiple_2009}. For example,  in network glasses the weakening is related to the local topological rearrangements of the atoms \cite{cai_1989, tsiok_1998, principi_2004, pal_2015} while for particle rafts it is attributed to the compression-induced changes in the interparticle interactions that evolve from  capillary bridge type to the frictional one \cite{varshney_amorphous_2012}. Similar crossover from  viscosity dominated lubrication regime to the frictional one has been a subject of interest in  dense granular flows and suspensions  \cite{langer_shear-transformation-zone_2008, varshney_amorphous_2012,trulsson_transition_2012,boyer_unifying_2011,lemaitre_what_2009}.
	

\end{document}